\documentclass[usenatbib]{mn2e}
\usepackage{graphicx}

\newcommand{\sigmat}{\sigma_\mathrm{T}}
\newcommand{\nh}{N_\mathrm{H}}
\newcommand{\me}{m_\mathrm{e}}
\newcommand{\mpr}{m_\mathrm{p}}
\newcommand{\ms}{M_\odot}
\newcommand{\tc}{T_\mathrm{C}}
\newcommand{\ef}{E_f}
\newcommand{\avs}{\langle F_E\rangle}
\newcommand{\avse}{\langle F_E(E)\rangle}
\newcommand{\fion}{F_\mathrm{ion}}
\newcommand{\lbol}{L_\mathrm{bol}}
\newcommand{\ledd}{L_\mathrm{Edd}}
\newcommand{\tstat}{T_\mathrm{stat}}
\newcommand{\mbh}{M_\mathrm{BH}}
\newcommand{\rc}{r_\mathrm{C}}
\newcommand{\taut}{\tau_\mathrm{T}}

\newcommand{\lesssim}{\mathrel{\hbox{\rlap{\hbox{\lower4pt\hbox{$\sim$}}}\hbox{$<$}}}}
\newcommand{\gtrsim}{\mathrel{\hbox{\rlap{\hbox{\lower4pt\hbox{$\sim$}}}\hbox{$>$}}}}
\newcommand{\beq}{\begin{equation}}
\newcommand{\eeq}{\end{equation}}
\newcommand{\beqa}{\begin{eqnarray}}
\newcommand{\eeqa}{\end{eqnarray}}

\title[Quasars: characteristic spectrum and radiative
heating]{Quasars: the characteristic spectrum and the induced
radiative heating}

\author[S. Yu. Sazonov et al.]{S. Yu. Sazonov$^{1,2}$\thanks{E-mail:
sazonov@mpa-garching.mpg.de}, J. P. Ostriker$^{3}$ and
R. A. Sunyaev$^{1,2}$\\
$^{1}$Max-Planck-Institut f\"ur Astrophysik,
Karl-Schwarzschild-Str. 1, 85740 Garching bei M\"unchen, Germany\\
$^{2}$Space Research Institute, Russian Academy of Sciences,
Profsoyuznaya 84/32, 117997 Moscow, Russia\\ 
$^3$Institute of Astronomy, Madingley Road, CB3 0HA Cambridge}

\begin{document}

\date{Accepted 2003 September 9. Received in original form 2003 May 14}


\maketitle

\label{firstpage}

\begin{abstract}
Using information on the cosmic X-ray background and the cumulative light of
active galactic nuclei (AGN) at infrared wavelengths, the estimated
local mass density of galactic massive black holes (MBHs) and
published AGN composite spectra in the optical, UV and X-ray, we
compute the characteristic angular-integrated, broad-band spectral
energy distribution of the average quasar in the universe. We
demonstrate that the radiation from such sources can photoionize and
Compton heat the plasma 
surrounding them up to an equilibrium Compton temperature ($\tc$) of
$2\times 10^7$~K. It is shown that circumnuclear obscuration cannot
significantly affect the net gas Compton heating and cooling rates, so
that the above $\tc$ value is approximately characteristic of both
obscured and unobscured quasars. This temperature is above typical gas
temperatures in elliptical galaxies and just above the virial
temperatures of giant ellipticals. The general results of this work
can be used for accurate calculations of the feedback effect of MBHs
on both their immediate environs and the more distant interstellar
medium of their host galaxies.
\end{abstract}


\section{Introduction}
\label{intro}

Massive black holes (MBHs) at cosmological distances came to our
attention more than three decades ago due to the enormous outflow from
them of high energy radiation. This observed fact, combined with the
inverse square law, implies that the immediate environs of these MBHs
undergo dramatic heating \citep{levsun71} when they are in the
luminous ``on'' state -- radiating at rates approaching Eddington
limit for their masses. Yet most work on active galactic nuclei (AGN)
or the environment of MBHs has ignored this ``feedback'' effect, even
though it is easy to demonstrate (e.g. \citealt*{cowetal78};
\citealt{parost01}) that it should dramatically alter both the
immediate environment of the MBH, from which accretion is occurring,
and the more distant interstellar medium of the galaxy within which
the MBH resides. The multiplicity of MHBs in the cores of galaxies
found in groups and clusters may even sensibly alter the entropy floor
of the ambient gas in these assemblages. 

But an accurate calculation of all of these effects could not be made
until we had accumulated and averaged the spectral output of
representative samples of AGN -- and that data has been missing; the
lack can now be remedied due to recent observational advances. Furthemore, some
means needed to be found to perform the averages, due to strong
variability of the sources with time, with viewing angle and from
object to object.

It is the purpose of this paper to address these issues quantitatively:
to compute the characteristic spectrum and all the important ``Compton
temperature'', $\tc$, of the average source and to estimate the consequences of
exposing gas of cosmic chemical composition to the radiation from such
sources. Specific applications of the general results of this paper
will be presented in future publications. Jumping ahead to our
conclusions, we constrain $\tc$ to a narrow range around $2\times
10^7$~K. This estimate is based on 1) measurements of the cumulative AGN
light at various wavelengths, 2) the measured local mass density of
MBHs and 3) published composite quasar spectra. Interestingly, our
estimate of the characteristic quasar Compton temperature is just above the
virial temperatures of giant elliptical galaxies and somewhat below
typical intracluster medium temperatures.

Before turning to our detailed calculation of quasar spectral output, it
may be useful to present some additional background evidence and
motivation. Comparison of the mass density of MBHs residing at the center of
nearby galaxies with the total radiation flux from all AGN suggests
that MBHs have grown by radiatively efficient accretion when their
host galaxies were quasars. Overall radiation efficiencies
[$\epsilon_\gamma\equiv \Delta E_\gamma/(\Delta m_\mathrm{BH}c^2)$] of
$\sim 0.1$--0.3 are estimated
\citep*{fabiwa99,saletal99,elvetal02,yutre02}. It is important to note
that this high mean efficiency is weighted by accretion rate; it
is possible that during most of the lifetime MBHs accrete much less
efficiently, but most of the mass must be accreted in (perhaps brief)
high efficiency episodes.

Luminous quasars were much more abundant in the past (at
redshifts $\sim 2$) than at the current epoch and one may reasonably
be puzzled as to why the most massive local ellipticals (such as M87),
which contain the most massive black holes, are so
quiescent. Elliptical galaxies do contain significant amounts of hot
($\sim 10^7$~K) gas capable of accreting onto central MBHs due to its
fast radiative cooling compared to the Hubble time. It 
has been suggested \citep{bintab95} that feedback
from the MBH may regulate accretion from the ambient gas, leading to an
oscillation-type behavior such that short periods of strong nuclear
activity are interchanged with much longer quiescence periods during which the
majority of observed galaxies are caught. This picture of brief
intervals of high efficiency accretion separated by long periods of
low level and low efficiency accretion is attractive in helping us 
understand the statistics of quasars and MBHs \citep{yutre02}. 

Continuing this theme, \citet{cioost97,cioost01} considered a
scenario in which the gas of a central cooling flow is heated, during a quasar
phase, above the galactic virial temperature by hard X-ray and gamma radiation
from the MBH, which leads to a degassing of the central regions of the
galaxy and switching off of the nucleus. A new episode of
nuclear activity can begin after a large amount of cooled gas has again
accumulated in the central regions of the galaxy. For this model to
work, the characteristic quasar Compton temperature must be higher than the
temperature of the cooling flow gas, otherwise the gas will be Compton
cooled by the low-frequency radiation from the nucleus instead of
being Compton heated. Regarding the Compton temperature as essentially
a free parameter, \citet{cioost01} presented solutions
for a number of $\tc$ values ranging from $5\times 10^7$ to $10^9$~K,
and the results are qualitatively independent of $\tc$ so long as
$c_{\rm C}^2=k\tc/\mpr>v^2_\ast$, the stellar velocity dispersion.

We note that the problem of gas preheating by X-rays
emergent from MBHs finds its analogy in other astrophysical situations:
near stellar-mass black holes located in globular clusters
\citep{ostetal76} as well as in X-ray binaries with stellar wind
accretion onto a black hole or a neutron star \citep{sunyaev78}.

\section{Basic assumptions and considerations}
\label{approach}

\subsection{Obscured vs. unobscured AGN}
\label{types}

In the standard AGN unification picture \citep{antonucci93}, the
active galactic nucleus is surrounded by an axisymmetric region
(hereafter called the torus) filled with cold and dense
material that intercepts and redistributes in wavelength and direction
a substantial fraction of the primary radiation. The source will be
classified as an unobscured (type 1) or obscured (type 2) AGN 
if its nucleus is observed directly -- at a small angle to the axis of
symmetry, or through the torus, respectively. One can also imagine
a different situation in which there are two physically different
populations of sources: naked AGN (type 1) and AGN enshrouded in
dense material (type 2). What plays the predominant role in nature,
orientation or the presence/absence of a dense envelope, is still a
matter of debate, but in either case type 1 and type 2 AGN exhibit
distinctly different spectral energy distributions when observed far
from the sources. 

It turns out that although circumnuclear absorption definitely
plays a crucial role in shaping the observed spectral energy 
distribution of quasar emission, its effect on the characteristic
Compton temperature is expected to be small. To demonstrate this let
us cast the usual definition of the Compton temperature
\citep[e.g.][]{levsun71} -- the plasma temperature at which net energy
exchange by Compton scattering between photons and electrons vanishes -- in the
following form: 
\beq
k\tc= 
\frac{1}{4}\frac{\int_0^{10\,\mathrm{keV}}EF_E\,dE+
\int_{10\,\mathrm{keV}}^{\infty} a(E)EF_E\,dE}
{\int_0^{10\,\mathrm{keV}} F_E\,dE
+\int_{10\,\mathrm{keV}}^{\infty} b(E) F_E\,dE}.
\label{tc1}
\eeq
Here $E$ is the photon energy, $F_E(E)$ is the radiation spectral flux
density, and the factors $a(E)$ and $b(E)$ represent Klein--Nishina
corrections that become of importance in the hard X-ray regime
(the corresponding explicit expressions are given in
Appendix). Note that equation (\ref{tc1}) is valid in the limit $k\tc\ll\me c^2
=511$~keV; it becomes inaccurate by more than 5\% at
$k\tc>10$~keV (in general the expression for Compton energy exchange
is nonlinear with respect to gas temperature, see e.g. \citealt{sazsun01}). 

As will be detailed in \S\ref{spectrum} and \S\ref{ctemp}, for the
characteristic spectral output of a type 1 quasar the Compton heating rate is
completely dominated by the high-energy integral
$\int_{10\,\mathrm{keV}}^{\infty} a(E)EF_E\,dE$ while most of the
Compton cooling is due to the component $\int_0^{10\,\mathrm{keV}} 
F_E\,dE$. The latter integral represents energy flux integrated over
the 'blue bump' and the infrared band. As a result, we may expect the estimate 
\beq
k\tc\approx\frac{1}{4}\frac{\int_{10\,\mathrm{keV}}^{\infty} a(E)EF_E\,dE}
{\int_0^{10\,\mathrm{keV}} F_E\,dE}
\label{tc2}
\eeq
to be accurate to a few per cent for type 1 AGN.

If we now consider lines of sight passing through an obscuring torus,
radiation emitted at $E\gtrsim 10$~keV will be little affected by
photoabsorption unless the torus is substantially Compton thick:
$\nh>$~a few $10^{24}$~cm$^{-2}$. Further, most of the radiation
emitted at $E\lesssim 10$~keV will be absorbed and reemitted in the infrared,
approximately retaining the total radiation flux due to energy
conservation. Thus, neither the numerator nor the denominator of the
RHS of equation (\ref{tc2}) is affected by obscuration to first order and
we may expect that equation (\ref{tc2}) with $F_E$ representing as before
the characteristic type 1 spectrum will also be a good approximation
for the characteristic Compton temperature of type 2 AGN. 
 
The above line of argument is directly applicable to the
scenario with two populations of sources, the unobscured and
the obscured. On the other hand, orientation-based unification
schemes assume that an obscuring torus is present in all AGN but it
covers a solid angle of less than $4\pi$. In the case of such
geometry, observers in unobscured directions will receive both direct
emission from the nucleus and a similar flux of reprocessed infrared
radiation from the torus. As a result, the characteristic Compton
temperatures of type 1 and type 2 AGN could be somewhat lower and
higher, respectively, than the angular averaged $\tc$ given by
equation~(\ref{tc2}).

We conclude that the characteristic Compton heating and cooling rates
(per particle) should be the same within a factor of $\sim 2$ in all
directions at a given distance from the AGN if the spectral
distribution of primary emission is isotropic. This implies that
knowledge of the angular averaged spectral output of a typical 
quasar would make it possible to calculate with good precision the
effect of Compton heating/cooling of gas around quasars. We note,
however, that other radiative mechanisms such as photoionization
heating and line cooling do depend strongly on the degree of
circumnuclear absorption (see \S\ref{ion}).

On the other hand, it is reasonable to expect that the two main spectral
components of the intrinsic AGN radiation, the blue bump and the hard
X-ray component, have somewhat different angular
distributions. This will of course influence the ratio of the heating and
cooling rates in equation (\ref{tc2}). Since one of the main
goals of this work is to obtain an unbiased estimate of the
characteristic Compton temperature of the average quasar, we will give
more weight in our computations to data on cumulative AGN light,
such as the cosmic X-ray background and the total infrared flux from
all AGN, than to another valuable source of information -- composite
spectra of type 1 quasars. The latter probably represent a
snapshot of quasar emission in a certain cone of angles or/and from a
certain population of objects.

\subsection{Cumulative AGN light}
\label{cumlight}

As mentioned above, our derivation of the spectral output of the
average quasar (in \S\ref{spectrum}) will be primarily based on
measurements of cumulative AGN light from the sky. It will be based on
the standard procedure described below.

Let $\epsilon_E(E,z)$ (erg~s$^{-1}$~cm$^{-3}$~keV$^{-1}$) be the angular
integrated spectral emissivity of AGN in a unit comoving
volume of the universe at redshift $z$. Then AGN located in the
redshift interval $[z,z+dz]$ will contribute to the locally measured
surface brightness of the sky the amount
\beq
dI_E(E)=\frac{c}{4\pi}\epsilon_E(E^\prime,z)\left|\frac{dt}{dz}\right|dz\,\,
({\rm erg~s}^{-1}~{\rm cm}^{-2}~{\rm sr}^{-1}~{\rm keV}^{-1}),
\label{dI}
\eeq
where $t$ is the cosmic time and $E^\prime=E(1+z)$ is the energy a
detected photon had when it was emitted. 

Integrating equation~(\ref{dI}) over redshift gives the intensity of
cumulative AGN light:
\beq
I_E(E)=\frac{c}{4\pi H_0}\int_0^\infty\frac{\epsilon_E((1+z)E,z)dz}
{(1+z)\left[\Omega_M(1+z)^3+\Omega_\Lambda\right]^{1/2}}.
\label{bgr}
\eeq
We adopt the concordance model cosmology \citep{ostste95}: $\Omega_M=0.3$ and
$\Omega_\Lambda=0.7$. Assuming that there is no spectral evolution of
AGN with redshift, we may further write
\beq
\epsilon_E(E,z)=\avse e(z).
\label{emis}
\eeq
Here, the function $e(z)$ describes the evolution with redshift of the
AGN comoving volume emissivity. 

Equations (\ref{bgr}) and (\ref{emis}) establish the relation between the 
measurable spectral energy distribution of the cumulative AGN light,
$I_E(E)$, and the angular integrated spectral output of the
average quasar, $\avse$, that we wish to determine and which
characterizes the global energy release via accretion onto MBHs.

The volume emissivity of type 1 quasars reaches a maximum at $z_{\rm
max}\sim 2$ \citep*[e.g.][]{schetal95}. Therefore, if the
cumulative spectrum $I_E(E)$ were known, one could roughly estimate
$\avse$ simply by shifting $I_E(E)$ to higher energies: $E\rightarrow
(1+z_{\rm max})E$. A more accurate treatment must be based on
equations (\ref{bgr}) and (\ref{emis}) and requires the knowledge of
the evolution function $e(z)$. We adopted the following evolution law
in our computations:
\beqa
e(z) &=&
\left\{
\begin{array}{ll}
(1+z)^3,\,\,& z\le z_0\\
e(z_0)\exp{(z_0-z)},\,\,& z>z_0
\end{array}
\right.
\nonumber\\
z_0 &=& 2.5.
\label{e_z1}
\eeqa
We have also considered the alternative law
\beq
e(z)=
\left\{
\begin{array}{ll}
(1+z)^3,\,\,& z\le z_0\\
e(z_0),\,\,& z>z_0,
\end{array}
\right.
\label{e_z2}
\eeq
We shall see (in \S\ref{high}) that both types of evolution lead to
practically the same results that are almost insensitive to the value
of the critical redshift if $z_0\gtrsim 2$. 

The above parametrization is motivated by the following data. The
evolution of type 1 quasars detected in the 2dF and Sloan Digital 
Sky optical surveys \citep{boyetal00,fanetal01} approximately matches
the cutoff pattern, equation~(\ref{e_z1}), with $z_0=2.5$. {\sl ROSAT}
soft X-ray (0.5--2~keV, \citealt*{miyetal00}) and {\sl BeppoSAX} hard
X-ray (2--10~keV, \citealt{lafetal02}) serveys suggest
that type 1 quasars evolve approximately following equation~(\ref{e_z2}) 
with $z_0=1.5$. The blazars observed at gamma-rays (above 
20~MeV) with {\sl CGRO}/EGRET evolve as $e(z)\propto (1+z)^3$ up to at least
$z\sim 2$, with no data available for higher redshift \citep{chimuk98}.

The most recent {\sl Chandra} deep X-ray (0.5--2~keV) and hard X-ray (2--8~keV)
observations have confirmed a decline in the number density of quasars
at $z>3$ but additionally revealed a second peak in
source counts at $z=0.5$--1 \citep{hasinger03,cowetal03}. This
low-redshift peak is, however, dominated 
by Seyfert galaxies, both unobscured and obscured, with 
X-ray luminosities in the range $L_X=10^{42}$--$10^{44}$~erg~s$^{-1}$,
as compared to the previously known $z\sim 2$~maximum in the number
density of quasars with $L_X\gtrsim 10^{44}$~erg~s$^{-1}$. For this reason, the
low-redshift peak is less prominent in terms of the contribution to
the total AGN X-ray light -- in fact about a third of the cosmic X-ray
background is now believed to come from AGN at $z<1$
\citep{baretal02}. This could move the redshift of a typical AGN 
contributing to the cumulative AGN light from $\langle z\rangle
\approx 1.5$ (see \S\ref{high}) to $\langle z\rangle \approx 1$ but
unlikely to lower redshift. Since the relevant redshift correction factor is
$1+\langle z\rangle$, the inferred average quasar spectrum could
change by some $25$\%. Considering this correction unimportant in view
of the remaining uncertainties on the observational side, we ignore the
$z\approx 0.7$ peak altogether in our calculations.

\section{The spectral output of the average quasar}
\label{spectrum}

In this section we construct the average quasar spectrum from its
pieces corresponding to different energy bands.
 
\subsection{High-energy emission}
\label{high}

\begin{figure}
\centering
\includegraphics[width=\columnwidth]{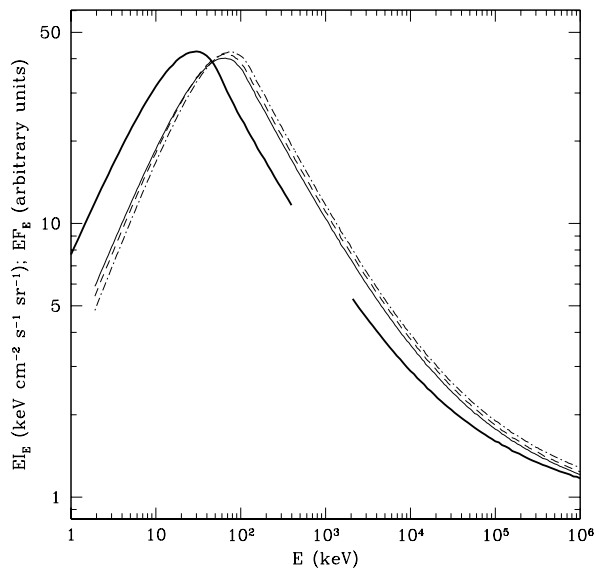}
\caption{Measured CXGB spectrum (heavy solid line with a gap) and the
adopted, rest-frame high-energy spectrum of the average quasar,
equation~(\ref{spec_xray}) (solid line). The redshifted light from a
population of such quasars distributed in redshift according to
equation~(\ref{e_z1}) has a spectral distribution that fits well the
CXGB spectrum. For comparison shown are two other source templates,
which generate similarly good fits to the CXGB spectrum when assuming
different evolution scenarios: equation~(\ref{e_z1}) with $z_0=3$
(dashed line) and equation~(\ref{e_z2}) with $z_0=2.5$
(dash-dotted line). 
}
\label{hard_bgr}
\end{figure}

We first consider emission above 2~keV. This spectral region is the main
contributor to the Compton heating by quasars. It is now well
established that the bulk of the cosmic X-ray and gamma-ray background
(hereafter CXGB), the observed spectrum of which is shown in
Fig.~\ref{hard_bgr}, is composed of contributions from AGN of varying
degree of obscuration, luminosity and radio loudness
\citep*{madetal94,cometal95,giletal01}. This conclusion is based on 1)
X-ray source counts, with some 80\% of the $E<5$~keV background having been
resolved into discrete sources \citep{hasetal01,rosetal02}, 2) data on
the hard X-ray spectra of Seyfert galaxies, 3) the measured
distribution of absorbing column densities in Seyferts
\citep*{risetal99}, 4) the redshift evolution of quasars and 5) for
the gamma-ray part tentatively also on the observed similarity of the
slopes of the CXGB spectrum and  the spectra of EGRET blazars
\citep{sreetal98}.

This allows us to identify, with some reservations as noted below (in
\S\ref{blazar} and \S\ref{nonagn}), the CXGB with the cumulative AGN
spectrum $I_E(E)$ and estimate the high-energy part of 
the average quasar spectrum $\avs$ using the equations of \S\ref{cumlight}. 

The CXGB spectrum was reliably measured in the 3--400~keV and
2~MeV--100~GeV bands by the {\sl HEAO-1} and {\sl CGRO} observatories, and a
useful formula fitting the measured intensities was 
proposed by \citet{gruetal99}. The spectrum peaks
(when plotted in units $EI_E$, see Fig.~\ref{hard_bgr}) at $\sim 30$~keV, is
characterized by an approximately constant slope ($I_E\propto
E^{-\alpha}$) $\alpha\approx 0.3$ at $\lesssim 20$~keV, by a varying
slope $\alpha\sim 1.5$ at energies above the peak up to $\sim 10$~MeV
and is approximately a power law with $\alpha\approx 1.1$ at higher energies.

Extrapolating the Gruber et al. formula for the CXGB spectrum down to
1~keV, we find $I_E(1~{\rm keV}) =8$~keV s$^{-1}$ cm$^{-2}$ sr$^{-1}$
keV$^{-1}$, which is 20\% below but is marginally consistent with the
newest estimate based on all-sky observations with RXTE/PCA
\citep{revetal03} and similarly below the value derived from a joint
analysis of {\sl ROSAT} and {\sl ASCA} observations of small celestial
fields \citep{miyetal98}, the latter estimate being affected by cosmic
variance. We thus apply the Gruber et al. fitting  
formula to describing the CXGB spectrum at all energies above 1~keV,
except for the poorly explored 0.4--2~MeV region which we 
omit from our analysis, and adopt a 20\% uncertainty in the
CXGB normalization.

We next adopt the template quasar spectrum
\beqa
\avs &=&
A\left\{
\begin{array}{ll}
E^{-\alpha}e^{-E/E_1},\,& 2~{\rm keV}\le E < E_0\\
B\left(1+kE^{\beta-\gamma}\right)E^{-\beta},\,& E\ge E_0
\equiv(\beta-\alpha)E_1
\end{array}
\right.
\nonumber\\
A &=& 2^\alpha e^{2/E_1},\,\,\,B=E_0^{\beta-\alpha}e^{-(\beta-\alpha)}
/(1+kE_0^{\beta-\gamma})
\nonumber\\
\alpha &=& 0.24,\,\, \beta=1.60,\,\, \gamma=1.06
\nonumber\\
E_1 &=& 83\,{\rm keV},\,\, k=4.1\times 10^{-3}. 
\label{spec_xray}
\eeqa
The spectrum $E\avs$ is shown in Fig.~\ref{hard_bgr}; it peaks at
approximately 60~keV. The convolution of this average spectrum with the
evolution law given by equation~(\ref{e_z1}) provides an excellent fit
to the CXGB spectrum. We note that the characteristic slopes of the
template (\ref{spec_xray}) in different spectral regions are
nearly the same as the corresponding original values for the CXGB
spectrum. This reflects the fact that only templates of the type
(\ref{spec_xray}) can lead to the observed CXGB spectrum; essentially only the
position of the peak is found from the fit. The high-energy template 
(\ref{spec_xray}) is normalized so that $\avs(2\,{\rm keV})=1$. Our
choice for the lower boundary of the spectrum (2~keV) is motivated by
the expectation that obscured sources contribute negligibly to the
average quasar spectrum at $\lesssim 2$~keV and also by the fact that
the high-energy component and the UV--soft X-ray component that will
be discussed in \S\ref{med} join near 2~keV in type 1 quasar spectra
(Laor et al. 1997).

Repeating the same fitting procedure for $z_0=3$ or assuming
the evolution given by equation~(\ref{e_z2}) with $z_0=2.5$, we
obtain average spectra that deviate from the previous one by less than
10\% (see Fig.~\ref{hard_bgr}). This demonstrates a very weak
dependence of the result on the assumed evolution scenario at $z>
2.5$, which reflects the fact that a typical quasar contributing to
our average spectrum is located at $\langle z\rangle=1.3$--1.8
(depending on $E$). Emission from substantially higher redshifts
contributes little to the background. 
 
\subsubsection{The average spectrum as a sum of the spectra of
obscured and unobscured sources}
\label{sum}

One can see that the average quasar spectrum just
derived differs notably below a few tens keV from observed X-ray
spectra of both unobscured and obscured AGN. This is, however,
consistent with this spectrum being the result of averaging over sources with a
range of line-of-sight photoabsorption columns. To illustrate this
point and also for the purposes of our further radiative energy
exchange computations, let us introduce two additional spectra of
which one will be representative of the spectra of type~1 quasars
and the other will characterize their obscured counterparts. We
impose the requirement that the appropriately weighted sum of these two
spectra should match well the globally averaged one. We exclude
from the current consideration blazars, assuming that they do not
contribute significantly to the average spectra below $\sim 500$~keV,
as will be justified in \S\ref{blazar}.

Most of the published information on AGN X-ray spectra pertains to
nearby ($z\ll 0.1$), relatively low-luminosity ($L_X=10^{42}$--$10^{44}$~erg
s$^{-1}$, the unabsorbed luminosity in the 2--10~keV band) Seyfert 1
and Seyfert 2 galaxies. \citet{peretal02} have recently
presented a sample of broad-band (0.1--200~keV) spectra obtained with
the BeppoSAX satellite for nine Seyfert 1s with $L_X$ ranging from
$5\times 10^{42}$ to $10^{44}$~erg s$^{-1}$. The measured spectral continua
above a few keV are well fitted by a model consisting of a power law
with high-energy exponential cutoff, $E^{-\alpha}\exp(-E/\ef)$, and a
Compton reflection component. Typical values for the power-law index
are $\alpha\approx 0.8$ and the cutoff energies $\ef$ span from about
70~keV to more than 300~keV, typically $\ef\sim 200$~keV. The relative
amplitude of the reflection component attributed to a weakly ionized
accretion disk is $R=0.5$--1. The above values of $\ef$ are
somewhat lower than the previous estimates ($\ef\gtrsim 250$~keV)
based on {\sl EXOSAT}, {\sl GINGA}, {\sl HEAO-1} and {\sl CGRO}/OSSE data
\citep{gonetal96}. We note that a rollover at $\sim$~50--100~keV was
first clearly detected in the spectrum of NGC~4151 with {\sl
GRANAT}/SIGMA \citep{jouetal92}.

A separate well-studied class of AGN is radio galaxies,
which can be considered the radio-loud counterparts of Seyfert
galaxies. Their X-ray spectra are similar to those of Seyferts, 
i.e. $\alpha\approx 0.8$, $\ef\gtrsim 100$~keV, except that the
reflection component seems to be weaker and is actually undetected in
most cases \citep*{wozetal98,eraetal00,graetal02}. Note that
the studied radio galaxies are on average more luminous ($L_X\sim$ a
few $10^{44}$~erg s$^{-1}$) than the Seyferts discussed above.

On the other hand, the available information on luminous ($L_X\gtrsim
10^{44}$~erg~s$^{-1}$) AGN located at $z> 0.1$, which usually are
classified as quasars, is rather scarce and generally limited to 
$E<20$~keV. However, the AGN X-ray luminosity function is
characterized by an increasing typical luminosity $L_\star$ with
redshift, so that $L_\star (z<1)\sim $~a few~$\times 10^{43}$~erg
s$^{-1}$ while $L_\star (z>1)\sim$~a few~$\times 10^{44}$~erg s$^{-1}$
\citep*{miyetal00,cowetal03}. As a consequence, a typical AGN
contributing to the CXGB is a quasar at $z\sim $1--2 with a luminosity
$L_X\gtrsim 10^{44}$~erg s$^{-1}$.

All of the available information on quasars (excluding blazars) is
consistent with their X-ray spectral continua being very similar to
those of local Seyfert and radio galaxies. In particular, the 
1--20~keV spectra of both radio-quiet and radio-loud type~1 quasars are
characterized by $\alpha\approx 0.8$
\citep*{lawtur97,reetur00,hasetal02}. In addition, high-energy spectral
cutoffs with $\ef\sim 100$~keV have been reported for at least two nearby 
($z\sim 0.1$) quasars, PG~1416--129 and MR~2251--178
\citep{stamai96,orretal01}. It appears that the Compton reflection
component is generally weak at least in the most luminous quasars with
$L_X\sim 10^{46}$~erg s$^{-1}$ \citep{reetur00}. 

\begin{figure}
\centering
\includegraphics[width=\columnwidth]{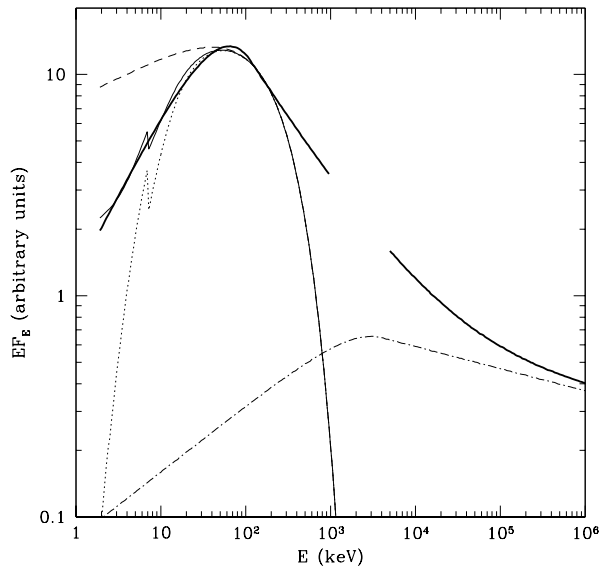}
\caption{{\sl Heavy solid line:} Adopted high-energy spectrum of the
average quasar, equation~(\ref{spec_xray}), with the gap at 1--5~MeV
approximately corresponding to the 0.4--2~MeV poorly
explored region in the CXGB spectrum (see Fig.~\ref{hard_bgr}). {\sl
Dashed line:} Adopted spectrum of the average unobscured quasar,
equation~(\ref{spec_xray_1}). {\sl Dotted line:} Adopted spectrum of the 
average obscured quasar, equations~(\ref{spec_xray_2}) and
(\ref{nh}). Note that above $\sim 50$~keV the obscured and unobscured
spectra coincide because photoabsorption becomes negligible. {\sl
Solid line:} Weighted sum, equation~(\ref{sum_12}), of the average obscured
and unobscured spectra. {\sl Dash-dotted line:} Estimated
upper limit, equation~(\ref{spec_xray_blazar}), on the contribution of
blazar emission to the average quasar spectrum.
}
\label{xray_spec}
\end{figure}

Summarizing the above facts, we adopt the following spectrum (plotted
in Fig.~\ref{xray_spec}) for the average unobscured quasar:
\beq
\avs ({\rm type 1})\propto E^{-0.8}e^{-E/200~{\rm keV}}.
\label{spec_xray_1}
\eeq

The chosen value for the cutoff energy not only falls comfortably in
the range of $\ef$ values found for nearby AGN and quasars,
but also leads to the right location of the peak in our globally
averaged quasar spectrum. We also note that the addition of a $R\sim
0.5$ reflection component to equation~(\ref{spec_xray_1}) would have a
fairly similar effect on the spectrum as slightly modifying the main
parameters -- $\alpha\approx 0.7$ and $\ef\approx 150$~keV. Since such
changes are within the current observational uncertainties and of the
same order as source-to-source variations, we ignore the reflection
component. We also ignore the usually observed fluorescence K$_\alpha$
iron line, since its equivalent width is small ($\sim 100$~eV, the
above references).

We next introduce another spectrum, which we attribute to the average
obscured quasar:
\beqa
\avs({\rm type 2})&=& \avs ({\rm type 1})
\nonumber\\
&&\times\int f(\nh)
\exp\left[-\sigma(E)\nh\right]\,d\nh,
\label{spec_xray_2}
\eeqa
where $\sigma(E)$ represents the photoelectric absorption cross section for 
solar chemical composition \citep{mormcc83}. We adopt a lognormal distribution
of absorption columns centered at $\nh=10^{24}$~cm$^{-2}$: 
\beq
f(\log \nh)=\frac{1}{\sqrt{\pi}}\exp\left[-(\log\nh-24)^2\right].
\label{nh}
\eeq
Our choice of the central value for $f(\log\nh)$ is in first place
dictated by the shape of the globally averaged quasar spectrum at
$E\lesssim 50$~keV, which is much flatter than the unabsorbed spectrum
(\ref{spec_xray_1}). In addition, the median value of the inferred 
$\nh$ distribution for the local population of Seyfert 2s is 
$\sim 10^{24}$~cm$^{-2}$ \citep{risetal99}. The
width of the $\nh$ distribution, equation~(\ref{nh}), was taken 
just sufficiently large to produce a relatively smooth
spectrum. For simplicity, we neglect here the Compton scattering of
nuclear radiation in the obscuring torus, which can significantly
modify the emergent spectrum when $\log\nh\gtrsim 24.5$ \citep*{matetal99}.

The absorbed spectrum given by equations~(\ref{spec_xray_2}) and
(\ref{nh}) is shown in Fig.~\ref{xray_spec}. We should mention that X-ray
spectra exhibiting a low-energy exponential cutoff have
been recently measured for several obscured quasars
\citep*{iwaetal01,steetal02,noretal02}.

Finally, we demonstrate in Fig.~\ref{xray_spec} that our
globally averaged quasar spectrum is well fit below $\sim 300$~keV
by a weighted sum of the adopted average spectra of type
1 and type 2 quasars:
\beq
\avs\approx 0.25 \avs({\rm type 1})+0.75 \avs({\rm type 2}).
\label{sum_12}
\eeq
This corresponds to the ratio $3:1$ of obscured to unobscured sources,
similar to that ($4:1$) estimated for the local population of Seyfert
galaxies \citep{mairie95}.

\subsubsection{Contribution from blazars}
\label{blazar}

Although there is little doubt that in the 2--300~keV band
our adopted average quasar spectrum represents mostly normal,
nonblazar quasar emission, the situation is more uncertain in the
gamma-ray region where blazars and non-AGN sources may contribute to
the average spectrum inferred from the CXGB.

According to a popular view \citep[e.g.][]{zdziarski96}, 
the CXGB above a few MeV is to a large part due to blazars, an AGN class
which we excluded from our consideration in \S\ref{sum}. The 
spectra of blazars are believed to be dominated by Doppler boosted 
radiation from a relativistic jet pointing close to our line of sight
\citep[e.g.][]{urrpad95}, as compared to the
quasi-isotropic emission from normal quasars. Regardless of whether
powerful jet emission is produced in most quasars but is usually
beamed away from us, or it occurs only in a small fraction of them,
obviously it is important to estimate the contribution of blazar
emission to the average quasar spectrum. As we shall demonstrate in
\S\ref{ctemp} using a constraint obtained below, blazar beamed emission is
at best a minor contributor to the global AGN energy output as well as
to the global radiative heating by AGN.

In contrast to normal quasars, it proves impossible to produce any
meaningful average spectrum for blazars, since blazar spectra 
vary dramatically from object to object and also with time: they 
usually consist of two broad peaks, whose positions and
relative amplitudes vary by orders of magnitude
\citep[e.g.][]{fosetal98}. Nonetheless, we can still derive an
interesting upper limit on the expected contribution from blazars to
the average quasar spectrum using the following facts:
\begin{enumerate}
\item
The 30~MeV--10~GeV spectra of some 50 blazars detected by
{\sl CGRO}/EGRET are consistent with being simple power laws, with the
average energy index $\alpha\approx 1.15\pm 0.04$ \citep{muketal97},
which is equal within the errors to the slope 
($\alpha\approx 1.10\pm 0.03$) of the CGRB spectrum in the
30~MeV--100~GeV range \citep{sreetal98}.  

\item
An estimated luminosity function of gamma-ray blazars and its
evolution, combined with the spectral information above, suggest that
blazars contribute at least a significant fraction of the CXGB in the
EGRET energy range \citep{chimuk98,sreetal98}. 

\item
The spectra of gamma-ray blazars peak at MeV energies
\citep{schoenfelder94,mcnetal95}. 

\item
The spectral slopes of gamma-ray blazars measured between 2~keV and a
few $\times 10^2$~keV vary considerably from object to object but 
tend to group around $\alpha\approx 0.7$ \citep{mcnetal95,kubetal98,tavetal02}.

\item
The contribution of all types of radio-loud quasars to the CXGB
background in the $\sim 1$~keV range is $\lesssim 10$\%
\citep{deletal94}, and the contribution of blazars, though very
uncertain, is likely several times lower, $\sim 1$\% \citep*{cometal96}.

\end{enumerate}

The following spectrum (see Fig.~\ref{xray_spec}) therefore provides a
plausible upper limit on the contribution of blazar emission to the
average quasar spectrum:
\beqa
\avs ({\rm blaz})=
\left\{
\begin{array}{ll}
0.082E^{-0.7}e^{-E/10^4\,{\rm keV}},\,& 2~{\rm keV}\le E < 4\,{\rm MeV}\\
1.52 E^{-1.1},\,& E\ge 4\,{\rm MeV}.
\end{array}
\right.
\label{spec_xray_blazar}
\eeqa
The spectrum (\ref{spec_xray_blazar})
was obtained by fixing the lower-energy and high-energy slopes at
$\alpha=0.7$ and $1.1$, respectively, and also assuming that the
blazar contribution to the globally averaged spectrum is 5\% at 2~keV
and $\approx 100$\% at $>100$~MeV. These conditions uniquely determine
the characteristic break energy.

\subsubsection{Contamination by other types of sources}
\label{nonagn}

The largest uncertainty is associated with the 300~keV--10~MeV part of the
average spectrum (see Fig.~\ref{xray_spec}). First, we
essentially do not know the intensities at 1--5~MeV (see
Fig.~\ref{xray_spec}) since the CXGB has not been reliably detected in the
corresponding 400~keV--2~MeV band (here we assume that a typical
source contributing to the CXGB is at $z\sim 1.5$). Moreover, other
types of astrophysical sources, not necessarily AGN, may provide the
major contribution in this energy range. The most promising candidate
proposed so far is Type 1a supernovae
\citep*{theetal93,ruietal01}. Finally, it cannot be ruled 
out that a subclass of blazars exhibiting a peculiar spectral bump at
$\sim 1$~MeV may provide a significant contribution
\citep{bloetal95,cometal96}. The last hypothesis remains highly
speculative at present, since only a few such ``MeV blazars'' have
been observed.

Fortunately, although the uncertainty associated with the average
quasar spectrum above 300~keV is large, it cannot affect significantly
our estimates of the global heating and cooling rates by AGN (see
\S\ref{ctemp}).

\subsection{Near-infrared to soft X-ray emission}
\label{med}

We next consider the 1~eV--2~keV band (which corresponds to
wavelengths $\lambda$ from $1.2$~$\mu$m to 6~\AA). Since there are
practically no data on cumulative AGN light for this spectral range,
our approach below will be different than in the preceding section.

In the standard AGN unification scenario, an obscuring torus of
dust and gas with $\nh\sim 10^{22}$--$10^{24}$~cm$^{-2}$ will
be transparent to X-rays above $\sim 10$~keV and also to infrared and
low-frequency radiation at $\lambda\gtrsim 10$~$\mu$m. All the
near-infrared to soft X-ray radiation emitted by the nucleus along
obscured lines of sight will be absorbed and reemitted at IR
and submillimeter wavelengths. We may thus expect that the spectrum of
the average quasar at 1~eV--2~keV will be completely dominated by
emission from unobscured sources. We can therefore take advantage of the ample
existing information on type 1 quasar spectra, which we may
condense in the simple spectral form
\beq
\avs=
1.20\left\{
\begin{array}{ll}
159 E^{-0.6} & 1~{\rm eV}\le E < 10~{\rm eV}\\
E^{-1.7}e^{E/2~{\rm keV}} & 10~{\rm eV} \le E < 2~{\rm keV}.
\end{array}
\right.
\label{spec_blue}
\eeq
Here 10~eV approximately corresponds to
$\lambda_{{\rm Ly}\alpha}=1216$~\AA\, and the adopted normalization is
such that the combined spectrum given by equations~(\ref{spec_xray})
and (\ref{spec_blue}) is continuous at 2~keV. 

In deriving equation~(\ref{spec_blue}), we took into account miscellaneous
recently published data on quasars, including the Hubble Space
Telescope catalog of UV spectra \citep{teletal02},
composite optical--UV spectra from the Sloan Digital Sky Survey (SDSS,
Vanden Berk et al. \citealt{beretal01}), 0.2--2~keV spectra from the
{\sl ROSAT} Bright Quasar Survey \citep{laoetal97}, and
statistics on the characteristic 2500~\AA--2~keV spectral slope,
$\alpha_{\rm OX}$ \citep*{yuaetal98,vigetal03}. We also used the atlas
of quasar energy distributions of \citet{elvetal94}. We note that the
combined contribution of resolved emission lines, including Ly$\alpha$, to the
total luminosity of the blue bump dominated by the continuum is small
\citep{teletal02}, but nevertheless an attempt has been made to take it
into account in the definition (\ref{spec_blue}). 

The $\alpha_{\rm OX}$ index is particularly relevant for our study, since it
together with the spectrum of the hard X-ray component determines the
ratio of the amplitudes of the hard X-ray and blue bumps in the quasar
spectrum. Equation (\ref{spec_blue}) yields $\alpha_{\rm OX}=1.4$. For
comparison, the \citet{elvetal94} mean radio-quiet and radio-loud
quasar spectra are characterized by $\alpha_{\rm OX}=1.4$ and
$\alpha_{\rm OX}=1.3$, respectively. From analysis of optical and {\sl
ROSAT}, {\sl Chandra} and {\sl XMM-Newton} X-ray data for optically
selected SDSS quasars (with redshifts up to 6), \citet{vigetal03} have
inferred a dependence of $\alpha_{\rm OX}$ on quasar rest-frame UV luminosity:
$\alpha_{\rm OX}\approx 1.5$ for $L_{2500~{\rm \AA}}\sim
10^{30}$~erg~s$^{-1}$~Hz$^{-1}$ and $\approx  1.7$ for $L_{2500~{\rm
\AA}}\sim 10^{32}$~erg~s$^{-1}$~Hz$^{-1}$. We note that these monochromatic
luminosities correspond to $L_{2-10\,{\rm keV}}$ ranging from $\sim
10^{44}$ to $\sim 10^{46}$~erg~s$^{-1}$ and, as was noted in
\S\ref{high}, the lower values from this range are characteristic of
the quasars producing the bulk of the global AGN energy output,
suggesting that $\alpha_{\rm OX}\approx 1.5$ might be a typical value
globally. 

We note that the template given by equation~(\ref{spec_blue}) possesses
a number of key properties of observed quasar spectra, including a turnover
near $\lambda_{{\rm Ly}\alpha}$ \citep{teletal02} and a gradual
flattening in the 0.2--2~keV band. The adopted template,
primarily inferred from recent UV observations, is redder than the 'big
blue bump' predicted by most theoretical accretion disk models whose
spectra extend to the extreme ultraviolet (see \citealt{korbla99} for
a review). We should also note that the spectra of radio-quiet and
radio-loud quasars are quite similar up to $\sim 100$~eV
\citep{elvetal94}, with $\approx 90$\% of all quasars being radio
quiet and this fraction being seemingly independent of redshift and
bolometric luminosity (\citealt{steetal00,iveetal02}; see however
\citealt{ciretal03}).

Given the remaining uncertainty in the $\alpha_{\rm OX}$
value characterizing the globally averaged quasar spectral output, we
estimate that the ratio of the optical--UV to high-energy integrated
radiation fluxes that follows from equations~(\ref{spec_xray}) and
(\ref{spec_blue}) for the average quasar spectrum is uncertain by
a factor of $\sim 2$.

We note that recent data seem to indicate that the scatter in
the near-IR to soft X-ray spectrum from quasar to quasar is fairly
small. Indeed, according to the Hubble Space Telescope observations
\citep{teletal02} the blue bump peaks invariably at 
$\lambda\approx 1200$~\AA~ (i.e. not far from 2500~\AA, the wavelength
appearing in the definition of $\alpha_{\rm OX}$) for quasars whose
monochromatic luminosities $\nu L_\nu$ at 1100~\AA~ range from a few
$10^{45}$ to a few $10^{47}$~erg~s$^{-1}$ (which corresponds to
$L_{2-10\,{\rm keV}}$ ranging from $\sim 10^{44}$ to $\sim
10^{46}$~erg~s$^{-1}$). On the other hand, the rms scatter in
$\alpha_{\rm OX}$ for a given luminosity is of the order of 0.1
\citep{vigetal03}, which corresponds to a factor of 2
scatter in the ratio $\nu L_\nu (2~{\rm keV})/\nu L_\nu (1200~{\rm
\AA})$, which is similar to our estimated uncertainty in this ratio
for the average spectrum.

\subsection{Medium-infrared to submillimeter emission}
\label{low}

Finally, we consider the spectral range $\lambda> 1~\mu$m, a major
contributor to the net Compton cooling. As for the high-energy region,
both obscured and unobscured sources are expected to significantly
contribute to the integrated AGN light at IR and submm
wavelengths. Although there is only fragmentary information on
$I_E(E)$ in this case, we shall use it below to derive some
constraints on the average quasar spectrum at $\lambda> 1~\mu$m. There
remains, after all available observations are utilized, 
a substantial uncertainty in the medium-IR luminosity of a typical
quasar. However, as will be noted and demonstrated below, we have an
integral constraint that provides an  additional strong bound. The
total luminosity of all quasars is limited by the observed,
low-redshift density of MBHs. This limits the integrated low-frequency
emission.

Our approach will be similar to several previous studies of
the contribution of AGN to the cosmic IR and submm background (hereafter 
CIB, \citealt*{almetal99,risetal02}). We take into account the
following data and limits related to the cumulative IR quasar light:

\begin{enumerate}

\item
The CIB spectrum (after subtraction of the cosmic microwave
background) at 125--2000~$\mu$m measured by the COBE/FIRAS experiment
\citep{fixetal98}.

\item
Luminous AGN, which produce the bulk of the cosmic X-ray background,
contribute less than $10$\% to the CIB background at 
850~$\mu$m, as inferred from the cross-correlation of SCUBA
submm sources with {\sl Chandra} deep X-ray surveys
\citep*{sevetal00,fabetal00}.

\item

The CIB at 15~$\mu$m: $E I_E(15~\mu{\rm m, total}) \approx
3$~nW~m$^{-2}$ sr$^{-1}$ \citep*{fraetal01,elbetal02}, and $\approx
17$\% of this intensity is due to AGN that make up $\sim 85$\% of the
2--10 keV background, as inferred from the cross-correlation of
{\sl XMM-Newton} and {\sl Chandra} deep X-ray surveys with ISOCAM
infrared surveys \citep*{fadetal02}. One can thus estimate the
cumulative quasar mid-IR light as $E I_E(15~\mu{\rm m})=0.5\pm
0.15$~nW~m$^{-2}$~sr$^{-1}$ \citep{fadetal02,elbetal02}. 

\item
Upper limits on the CIB at $\lambda<40~\mu$m set by TeV cosmic opacity
measurements toward blazars \citep{fraetal01}. 

\end{enumerate}

\begin{figure}
\centering
\includegraphics[width=\columnwidth]{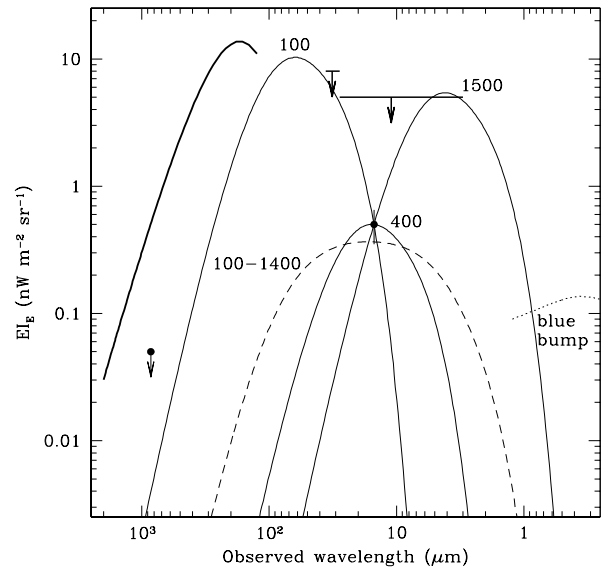}
\caption{Observational constraints on the spectrum of cumulative AGN
IR and submm light: the data point with error bars at 15~$\mu$m and
the upper limit at 850~$\mu$m. Also shown are the COBE/FIRAS spectrum of 
the total extragalactic background (CIB) at 125--2000~$\mu$m (thick
solid line) and upper limits on the CIB at 3--40~$\mu$m. See text for
references to the data. All these data cobined allow a large variety of
possible IR spectra of the average quasar, including the modified
black body emission spectrum, equation~(\ref{spec_dust}); the result
of convolution of this model spectrum with the quasar evolution function,
equations~(\ref{bgr})--(\ref{e_z1}), is shown for $\beta=2$ and three
values of the dust effective temperature (as indicated in K above the
corresponding curves) that either maximize or minimize the
integral $E I_E$. The additional integral constraint given by
equation~(\ref{ir_flux}) leads to the adopted average quasar IR
spectrum, equation~(\ref{spec_ir}) (dashed line, after
convolution with the quasar evolution function). Also shown (dotted
line) is the adopted average quasar spectrum in the 
blue bump region, equation~(\ref{spec_blue}), after convolution with the quasar
evolution function, with the normalization fixed by the cosmic X-ray
background.
}
\label{ir_bgr}
\end{figure}

The above data are displayed in Fig.~\ref{ir_bgr}. We also show 
the expected cumulative quasar light spectrum in the near-infrared ($\lambda<
1$~$\mu$m), calculated from the blue bump template,
equation~(\ref{spec_blue}), with the normalization fixed by the cosmic
X-ray background. Evidently only loose constraints can be obtained on
the far-IR emission of the average quasar from the presented data
alone. Fortunately, the radiative energy exchange between the quasar and the
surrounding gas does not depend at all on the spectral shape at
IR wavelengths. The only important thing is the total luminosity emitted
in this spectral band, and it is possible to obtain some constraints
on this quantity from the observational data presented in
Fig.~\ref{ir_bgr}.

Indeed, it is widely accepted that the mechanism responsible for the
IR emission of radio-quiet quasars (and partly for that of radio-loud
quasars) is thermal radiation from radiatively heated dust
\citep{barvainis87}. Therefore, whichever is the spectral 
distribution of the IR emission from a quasar or from an ensemble of
quasars, it cannot be narrower than that of modified black body
emission from monotemperature dust:
\beq
F_E ({\rm dust})\propto\frac{E^{3+\beta}}{\exp(E/kT_{\rm d})-1},
\label{spec_dust}
\eeq
where $T_{\rm d}$ is the dust temperature, $\beta\sim 1$--2 is the
emissivity index, and we have assumed the emission to be optically
thin [the effect of a significant optical depth can be mimicked by
varying the values of $\beta$ and $T_{\rm d}$ in equation~(\ref{spec_dust})].

We show in Fig.~\ref{ir_bgr} several examples of spectra of the cumulative
quasar light computed from equations~(\ref{bgr})--(\ref{e_z1}) for the
template spectrum (\ref{spec_dust}) taking different $T_{\rm d}$
values. Normalizing the resulting spectra at 15~$\mu$m to
0.5~nW~m$^{-2}$~sr$^{-1}$, the current best estimate for $EI_E$, we can
derive an upper limit on the integrated IR emission from quasars as a 
function of the characteristic dust temperature $T_{\rm d}$. In this
way we find that $\int I_E\,dE< 15$~nW~m$^{-2}$~sr$^{-1}$
and $< 8$~nW~m$^{-2}$~sr$^{-1}$ for $T_{\rm d}=100$ and 1500~K,
respectively. These limiting dust temperatures lead to 
maximal luminosities. The more general upper limit
$\int I_E\,dE \lesssim 20$~nW~m$^{-2}$~sr$^{-1}$ is applicable to any
realistic spectral distribution, characterized by a range of $T_{\rm d}$.

On the other hand, there is a more stringent, lower limit on the
cumulative, frequency-integrated far-IR light from quasars: 
\beq
I_{\rm IR}\equiv \int_{0}^{1{\rm eV}} I_E\, dE> 0.8\pm
0.3\,\,{\rm nW}~{\rm m}^{-2}~{\rm sr}^{-1}. 
\label{ir_flux_low}
\eeq
This minimum corresponds to dust emission with $T_{\rm d}\approx
400$~K (see Fig.~\ref{ir_bgr}), and the quoted error results 
from the uncertainty in the observed flux of cumulative AGN light at
$15\, \mu{\rm m}$. 

Let us now derive an upper limit on $I_{\rm IR}$, following the
argument of \citet{soltan82}. The mass density of MBHs in the
local universe is estimated as $\rho_{\rm BH}=(3\pm 1)\times 10^5
M_\odot$~Mpc$^{-3}$ \citep{yutre02,allric02}, taking $h=0.72\pm
0.05$ \citep{speetal03}. Assuming that all of this mass 
has been accumulated  through accretion with a redshift-independent
mass-to-radiation conversion efficiency $\epsilon_\gamma$ gives an
upper limit on the total, frequency-integrated radiation flux from all quasars:
\beqa
I_{\rm total}\equiv \int_0^\infty I_E\, dE
 &\le & \frac{\epsilon_\gamma c^3\rho_{\rm BH}}{4\pi}
\frac{\int e(z)/(1+z)|dt/dz|dz}{\int e(z)|dt/dz|dz}
\nonumber\\
&\approx& (1.7\pm 0.6)\frac{\epsilon_\gamma}{0.1}\,\,
{\rm nW}~{\rm m}^{-2}~{\rm sr}^{-1},
\label{bl_flux}
\eeqa
where we have again used equation~(\ref{e_z1}) to describe
the quasar emissivity evolution (although the dependence of the result
on $e(z)$ is weak). 

Now, the total flux $I_{\rm total}$ consists of contributions from the
high-energy region ($>1$~keV), the blue-bump region (1~eV~$<E<1$~keV)
and the low-frequency region ($<1$~eV). The first of these components
can be found directly by integrating the CXGB spectrum shown in Fig.~\ref{hard_bgr}:
\beq
I_{\rm X-ray}= 0.29\pm 0.06\,\,{\rm nW}~{\rm m}^{-2}~{\rm sr}^{-1},
\label{xg_flux}
\eeq
where the quoted error has been propagated from the uncertainty in
the CXGB normalization.

A more conservative estimate can be obtained by integrating the X-ray
background over the 1--150~keV range, where it is certainly dominated
by AGN emission:
\beq
I_{1-150\,{\rm keV}}= 0.21\pm 0.04\,\,{\rm nW}~{\rm m}^{-2}~{\rm sr}^{-1},
\label{x_flux}
\eeq

The corresponding integral over the blue bump can be found by substituting
the template given by equations~(\ref{spec_xray}) and
(\ref{spec_blue}) into equations~(\ref{bgr})--(\ref{e_z1}), and using the
normalization provided by the CXGB:
\beq
I_{\rm blue}=0.4_{-0.2}^{+0.4}\,\,{\rm nW}~{\rm m}^{-2}~{\rm sr}^{-1},
\label{blue_flux}
\eeq
where we have roughly estimated the observational uncertainty in the
average optical to soft-X-ray spectrum of type 1 quasars. 

We can now estimate the maximum possible integrated quasar infrared ($\lambda
>1$~mm) light:
\beqa
I_{\rm IR} &=& I_{\rm total}-I_{\rm blue}-I_{1-150\,{\rm keV}}
\nonumber\\
&\le & (1.7\pm 0.6) \frac{\epsilon_\gamma}{0.1}-0.7\pm 0.3
\,\,{\rm nW}~{\rm m}^{-2}~{\rm sr}^{-1}.
\label{ir_flux_up}
\eeqa

One can see that the upper limit on $I_{\rm IR}$,
equation~(\ref{ir_flux_up}), is consistent with the lower limit on $I_{\rm
IR}$, equation~(\ref{ir_flux_low}), if $\epsilon_\gamma$ is above 
0.05, implying that MBHs have grown by
radiatively efficient accretion via a standard disk. A similar
conclusion was earlier reached by other researchers
\citep[e.g.][]{elvetal02,yutre02}. If we additionally
require that $\epsilon_\gamma\le 0.1$, then we find from the limits given by
equations~(\ref{ir_flux_low}) and (\ref{ir_flux_up}) that
\beq
I_{\rm IR}=(1.0\pm 0.6)\,\,{\rm nW}~{\rm m}^{-2}~{\rm sr}^{-1}.
\label{ir_flux}
\eeq

It was implicit in the above treatment that the radiative output
of quasars is purely the result of accretion onto MBHs. Recent observational
data lend support to this view, seemingly excluding starburst activity as the
dominant contributor to the bolometric luminosity of powerful AGN
producing the bulk of the cosmic X-ray background. Indeed,
high-quality spectra obtained with the {\sl ISO} satellite for
dozens of optically bright quasars \citep{haaetal00,poletal00,andetal03},
plotted in units $EF_E$, exhibit a broad bump rising from 1--2~$\mu$m 
(followed by the blue bump on the short-wavelength side) and declining
in most cases at $\lambda\gtrsim 60$~$\mu$m. This bump is interpreted
as multitemperature thermal emission from warm dust, with the maximum
temperature (1000--1500~K)  representing the evaporation temperature
of dust grains. In addition, the far-IR spectra measured for several
obscured quasars also tend to peak below 30~$\mu$m
\citep[e.g.][]{baretal95,deatre01}.

The fact that the bulk of the infrared luminosity of quasars is 
emitted at wavelengths shorter than $60$~$\mu$m implies that dust cooler
than $\sim 50$~K is typically not energetically important. This, according
to the standard theory \citep[e.g.][]{rowan95}, suggests
that the observed infrared emission is mostly the result of
reprocessing by dust of the optical--UV radiation from 
the nucleus, rather than resulting from starburst activity.

Overall, the available data are consistent with the dominant
contribution to the infrared luminosity of a typical quasar being due
to dust heated to 50--1500~K by the radiation released 
as a result of accretion onto the central MBH. For this reason we
consider equation~(\ref{ir_flux}) a robust estimate that can be used to
normalize the average IR quasar spectrum, for which we adopt the
following template:
\beq
\avs= 1.5\times 10^{42}\int_{100~{\rm K}}^{1400~{\rm K}}
\frac{E^5 T_{\rm d}^{-7}}{\exp(E/kT_{\rm d})-1}\,dT_{\rm d},\,\, E<
1~{\rm eV},
\label{spec_ir}
\eeq
where $E$ and $T_{\rm d}$ are measured in keV and K, respectively.

This spectrum, which bears a greater resemblance to observed quasar spectra
than equation~(\ref{spec_dust}), represents optically
thin modified ($\beta=2$) black-body emission of dust characterized by
a range of temperatures (100--1400~K) and constant emitted luminosity per unit 
logarithmic $T_{\rm d}$ interval (so that the spectrum has a flat core
when plotted in $EF_E$ units). The numerical coefficient in
equation~(\ref{spec_ir}), which is defined relative to the X-ray component,
equation~(\ref{spec_xray}), enables that $\int_0^{1{\rm eV}} I_E\,dE=
1.0$~nW~m$^{-2}$~sr$^{-1}$. The adopted upper $T_{\rm d}$ value 
enables continuity at 1~eV between the distributions 
given by equations~(\ref{spec_ir}) and (\ref{spec_blue}), while the
lower $T_{\rm d}$ boundary and the spectral form have been chosen
rather arbitrarily to allow the predicted cumulative quasar
light to be consistent with the data point  at 15~$\mu$m (see
Fig.~\ref{ir_bgr}).

\subsection{The broad-band spectrum}
\label{broad}

\begin{figure}
\centering
\includegraphics[width=\columnwidth]{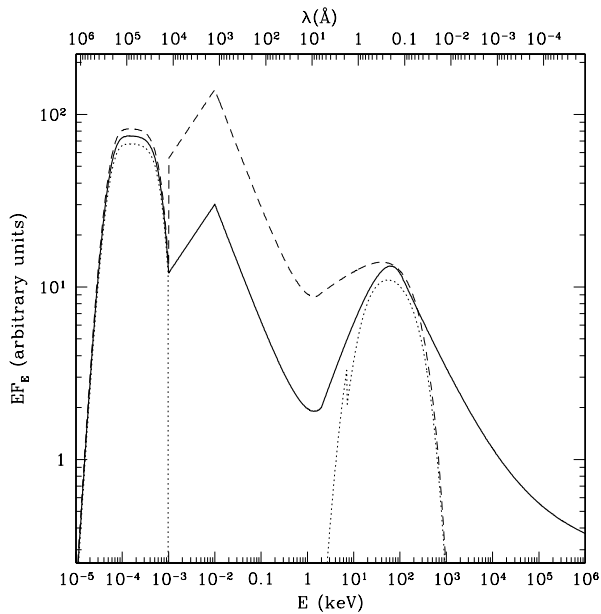}
\caption{{\sl Solid line:} Broad-band spectrum of the average quasar
adopted in this paper. {\sl Dashed line:} Adopted
spectrum of the average unobscured quasar, rescaled by a factor of 1.1
for visibility. {\sl Dotted line:} Adopted spectrum of the average
obscured quasar, rescaled by a factor of 0.9. The weighted sum of the
adopted type 1 and type 2 spectra (not shown) matches well the average
spectrum below 300~keV.
}
\label{broad_spectrum}
\end{figure}

We are now finally in a position to put together the different pieces
of the average quasar spectrum. The resulting broad-band
spectrum is shown in Fig.~\ref{broad_spectrum}. It has three broad
maxima in terms of energy output: one in the medium infrared, the blue
bump and another one at $\sim 60$~keV. It is necessary to note that the radio
($\lambda>1$~cm) component of the quasar spectrum is 
unimportant, because its contribution to the total energy output and
consequently to the net Compton cooling rate is very small. This
follows from the estimated surface brightness of the extragalactic radio
background -- $\int I_E\,dE\sim 
10^{-4}$~nW~m$^{-2}$~sr$^{-1}$ \citep{bridle67,dwebar02} and also from
spectroscopy of individual quasars -- the radio luminosity of
radio-quiet and radio-loud quasars is typically 3 and 5 orders of magnitude
smaller, respectively, than the luminosity of the IR or blue bumps
\citep[e.g.][]{elvetal94}. We note, however, that in the
vicinity of parsec-scale quasar jets simulated Compton heating by
their radio radiation can become important \citep{levsun71}.

Since we shall study in the following sections the effect of AGN
obscuration on gas heating and cooling, we show in
Fig.~\ref{broad_spectrum} two additional spectra that may be 
considered representative of type 1 and type 2 quasars. We have
defined these templates in a somewhat arbitrary manner, which is
unavoidable for several reasons, in particular the distribution of emitted
energy between the hard X-ray component and the blue bump may be
intrinsically anisotropic, i.e. depend on the line of sight even in the
absence of circumnuclear absorption; also the obscuration pattern may be
more complicated than assumed, e.g. some quasars may appear obscured
in the UV but unabsorbed in the X-ray, etc. Specifically we
required that
\begin{itemize} 

\item
the ratio of obscured to unobscured quasars be 3 to 1,

\item
in the X-ray region, the type 1 and type 2 spectra be given by
equation~(\ref{spec_xray_1}) and equation~(\ref{spec_xray_2}), respectively,

\item
the type 2 spectrum have no emission in the blue bump region,

\item
the infrared bump be the same for all three considered templates.  

\end{itemize}
The weighted sum of the adopted obscured and unobscured
quasar spectra matches well the globally averaged quasar spectrum below
$\sim 300$~keV. We recall that this upper boundary results directly
from the cutoff energy ($200$~keV) that we adopted in equation
(\ref{spec_xray_1}) and that there remains a significant observational
uncertainty in the position of this cutoff for Seyfert galaxies and
quasars.   

\section{Compton temperature}
\label{ctemp}

We can now determine the Compton temperature $\tc$ characterizing   
the spectral output of the average quasar, which is the gas temperature
at which Compton heating balances Compton cooling. An accurate
result in the limit $k\tc\ll \me c^2=511$~keV can be obtained
from equations given in Appendix by requiring that
$(dW/dt)_{+}+(dW/dt)_{-}=0$. 

Let us note here that if additionally $E\ll 
\me c^2$, the Compton cooling rate becomes proportional to the
frequency-integrated flux $\int \avs dE$. We may thus expect that for
our adopted average quasar spectrum most of the cooling will be
provided by the infrared component, with additional smaller  
contributions coming from the optical--UV and high-energy
bands. Klein--Nishina corrections act to further reduce the cooling by
hard X-rays and gamma-rays. On the other hand, the Compton heating
rate is, in the limit $E\ll \me c^2$, proportional to another integral,
$\int E\avs dE$, which is completely dominated by the high-energy
component of the quasar spectrum. However, also in this case
Klein--Nishina corrections dramatically diminish the heating by
photons with $E\gtrsim 100$~keV.

Both the Compton heating and the Compton cooling integrals are
well constrained for the adopted average quasar spectrum. First, the
spectral shape of the high-energy component is known very well at
least up to 300~keV and its normalization is uncertain to within $\sim
20$\% -- see equation (\ref{xg_flux}). Second, the uncertainty in the
radiation flux integrated over the entire average spectrum is $\sim
40$\%, as follows from equations (\ref{ir_flux_low}), (\ref{bl_flux})
and (\ref{ir_flux_up}). We can thus determine the characteristic
Compton temperature quite accurately:
\beq
\tc= (1.9\pm 0.8)\times 10^7\,{\rm K}.
\label{ctemp_mean}
\eeq

We emphasize that the surprisingly small uncertainty in the final
result has resulted from three facts: 1) good knowledge of both the
spectrum and normalization of the CXGB, 2) the lower limit on the
cumulative AGN light in the infrared band that follows from the
cross-correlation of deep X-ray and infrared surveys and 3) the upper
limit on the total flux from all quasars inferred from the local density of
MBHs. It is interesting that the detailed properties of the blue bump play
a minor role in obtaining equation (\ref{ctemp_mean}). Indeed, the
contribution of the blue bump to the Compton heating rate is negiligible,
while its contribution to the Compton cooling rate is of the order of
25\%, which is less than the uncertainty in the $\tc$ value for the
average quasar. 

The value given by equation (\ref{ctemp_mean}) corresponds to the
entire spectrum extending into the gamma-ray band. If we exclude from
consideration the spectral segment above 300~keV whose AGN origin
still needs to be proven or disproven, we find 
\beq
\tc (<300~{\rm keV})= 1.3\times 10^7\,{\rm K}.
\label{ctemp_500}
\eeq
Therefore, AGN emission above 300~keV contributes less
than $30$\% to the total Compton heating rate (its contribution to the
Compton cooling rate being negligible). The contribution of emission
above 500~keV is $<20$\% and this excludes blazar beamed emission as a
major contributor to the global heating due to AGN. 

We can similarly calculate the Compton temperatures corresponding 
to the adopted type 1 and type 2 quasar spectra (see
Fig.~\ref{broad_spectrum}):
\beq
\tc ({\rm unobsc})= 0.8\times 10^7\,{\rm K},
\label{ctemp_type1}
\eeq
\beq
\tc ({\rm obsc})= 2.0\times 10^7\,{\rm K}.
\label{ctemp_type2}
\eeq
Because we have taken substantial freedom in defining the type 1
and type 2 spectra, the above values are clearly more uncertain than the
Compton temperature corresponding to the globally averaged spectrum.

We should note that the above calculation of Compton energy exchange
is valid in the case of a  Maxwellian momentum distribution of
electrons, and thus assumes that the time scale for isotropization of
electron momenta by mutual collisions is shorter than that for
scattering of hard photons by electrons. We also explicitly assumed in
our calculations that there is no significant bulk motion of the
plasma caused by the gravity and radiation pressure of the central
MBH. Both of these assumptions can be violated very close to quasar nuclei.

\section{Effects of exposing gas to the radiation from the average quasar}
\label{effects}

Having built the spectrum and determined the characteristic Compton
temperature of the average quasar, we can now estimate the
consequences of exposing gas of cosmic chemical composition to the
radiation from such sources.

\subsection{Maximum distance of heating in the low density limit}
\label{rmax}

As was noted in \S\ref{intro}, it is well established that
MBHs sitting in the centers of local galaxies have experienced an
epoch or multiple epochs of rapid growth accompanied by the appearance of a
quasar. One may therefore ask: to which maximum radius could a MBH
currently of mass $\mbh$ heat during its life time gas from a low
initial temperature to the quasar Compton temperature? Such formulation
assumes that the gas is of sufficicnetly low density that it is fully
photoionized so that only Compton heating and cooling are important. 

For gas of temperature $T$ the heating rate per electron at a distance
$r$ from the quasar is
\beq
\frac{dW}{dt}=\frac{L(t)\sigmat}{4\pi r^2}\frac{4k(\tc-T)}{\me c^2},
\eeq
where $L$ is the quasar luminosity, and we have assumed that
the source is isotropic. Since the quasar is powered by accretion, the
luminosity can be related to the rate of growth of the MBH:
\beq
L(t)=\epsilon_\gamma\frac{d\mbh}{dt}(t) c^2.
\eeq
Hence, the total energy received during the growth of the MBH by an
electron--proton pair located at $r$ is 
\beq
\Delta W=\frac{k\tc}{\me c^2}\frac{\sigmat}{\pi r^2}\epsilon_\gamma\mbh c^2,
\eeq
assuming that $T\ll\tc$.

If we now require that each electron--proton pair receive an
energy at least $\Delta W=3k\tc$, we find the maximum distance out to
which this can be done:
\beq
\rc=\left(\frac{\sigmat\epsilon_\gamma\mbh}{2\pi\me}\right)^{1/2}
=0.4~{\rm kpc} \left(\frac{\epsilon_\gamma}{0.1}\right)^{1/2}
\left(\frac{\mbh}{10^8\ms}\right)^{1/2}.
\label{rc}
\eeq

\subsection{Heating/cooling of a partially ionized gas}
\label{ion}

Up to now our attention has been focused on Compton heating and
cooling. However, gas exposed to intense quasar radiation 
may be sufficiently dense to remain only partially ionized. Photoionization
heating as well as cooling through continuum and line emission will
then be important. We consider this situation below. 

We assume that the gas is optically thin and is in ionization
equilibrium. Note that the characteristic ionization and
recombination times are typically much shorter than the time scale for
Compton heating/cooling. We adopt the solar element abundances from
\citet*{greetal96}, unless stated otherwise. Under these assumptions,
the ionization balance as well as the heating and cooling rates are
fully determined by the instantaneous gas temperature, the radiation
spectral distribution and the ionization parameter \citep*{taretal69},
which we define here as
\beq
\xi\equiv\frac{\lbol}{nr^2}=1.4\times 10^9 
\frac{\lbol}{\ledd}\frac{\mbh}{10^8\ms}
\frac{1~{\rm cm}^{-3}}{n}\left(\frac{1~{\rm pc}}{r}\right)^2,
\label{xi}
\eeq
where $\lbol$ (erg~s$^{-1}$) is the bolometric, angular-integrated
luminosity of the central source, $\ledd$ is the Eddington luminosity
for a given MBH mass, $n$ the hydrogen nucleus density and $r$ the
distance from the source. To proceed we assume that $3/4$ of the whole
sky as seen from the MBH is obscured while $1/4$ is clear and that the
emergent spectrum is described by either the adopted type 2 or type 1
spectrum (see Fig.~\ref{broad_spectrum}), dependent on from which part of the sky the source is observed. According to this picture, observers located
at the same distance but in different directions from the source will
receive equal radiation fluxes above $\sim 20$~keV and below $\sim
1$~eV, but those looking through the obscuring torus will not see any
optical, UV or soft X-ray radiation.

\begin{figure}
\centering
\includegraphics[bb=90 140 450 700, width=\columnwidth]{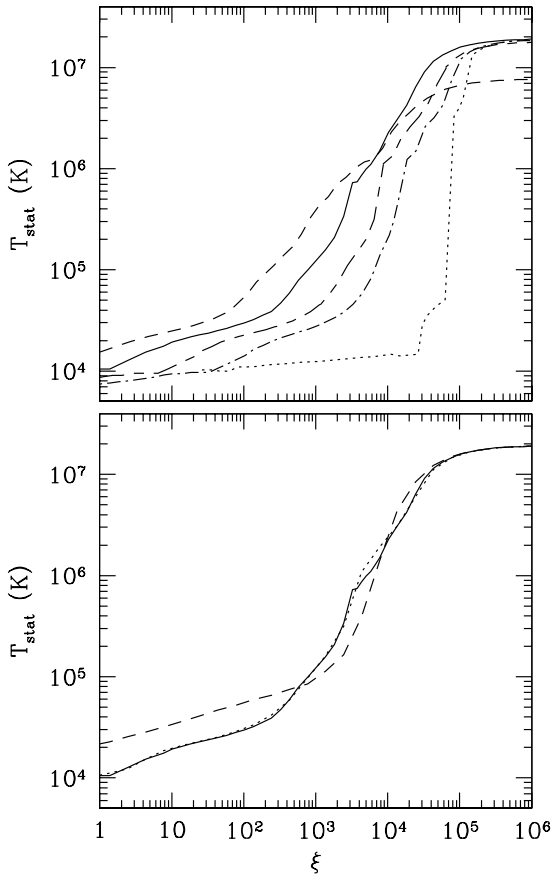}
\caption{{\bf (a)} Dependence on the ionization parameter,
equation~(\ref{xi}), of the stationary temperature of plasma of solar
chemical composition exposed to the different radiation spectral
distributions considered in this paper: the globally averaged quasar
spectrum (solid line), type 1 spectrum (dashed line), type
2 spectrum (dotted line), and additionally $\langle F_E\rangle ({\rm type
2})+0.01\langle F_E\rangle ({\rm type 1})$ (dot-dashed line) and
$\langle F_E\rangle ({\rm type 2})+0.05\langle F_E\rangle ({\rm type
1})$ (short-dash-long-dashed line). {\bf (b)} The effect of plasma
chemical composition on the $\tstat(\xi)$ curve for the average quasar
spectrum. {\sl Solid line:} solar element abundances, {\sl dotted
line:} iron is absent, {\sl dashed line:}
H--He gas.
}
\label{xi_temp}
\end{figure}

We performed computations using the latest version of XSTAR
\citep{kallman02}, to which we have added a block responsible for
calculating Compton heating and cooling from equations~(\ref{cheat})
and (\ref{ccool}). Our code also accounts for an additional heating 
caused by Compton scattering of hard X-rays off electrons
bound in hydrogen and helium atoms. To this end we use the simple
approximation that the Compton heating rate per electron is the same
for bound and free electrons and is described by
equation~(\ref{cheat}). This implies that the net Compton heating rate
per unit volume does not depend on the fraction of free 
electrons in the plasma, as opposed to the Compton cooling rate, which
is proportional to this fraction. This has been shown to
be a good approximation for photons with energies above a few~keV
\citep*{basetal74,sunchu96}, and such photons completely dominate the
Compton heating in our case. 

Since the definition of $\xi$ adopted in XSTAR is
different from ours, equation~(\ref{xi}), we provide here the
relations between $\lbol$ and the ionizing flux $\fion$ (between
13.6~eV and 13.6~keV) in obscured and unobscured directions as well as
its average over the sky:
\beqa 
\frac{\langle 4\pi r^2 \fion\rangle}{\lbol} &=& 0.13, 
\nonumber\\
\frac{4\pi r^2\fion}{\lbol} ({\rm unobsc}) &=& 0.50, 
\frac{4\pi r^2\fion}{\lbol} ({\rm obsc})=0.012.
\eeqa

Suppose now that the irradiated gas has had enough time to achieve 
thermal equilibrium. Fig.~\ref{xi_temp}a shows the gas stationary
temperature $\tstat$ as a function of $\xi$ for each of the different
spectral templates presented in
Fig.~\ref{broad_spectrum}. Fig.~\ref{xi_temp}b illustrates the effect
of gas enrichment by metals and iron on the $\tstat(\xi)$ curve.  
We see that the gas reaches the Compton temperature given by
equations~(\ref{ctemp_mean})--(\ref{ctemp_type2}) when $\xi\gtrsim
10^5$. In this region the gas is fully photoionized and the thermal
balance is dominated by Compton heating and cooling. At 
$\xi\lesssim 10^5$, the $\tstat(\xi)$ curves are fairly similar for
the globally averaged spectrum and for the type 1 spectrum,
and are mainly determined by the balance between photoionization heating and
various cooling mechanisms (see \citealt{kalmcc82} for a detailed
discussion of the underlying physics).

Therefore, for given luminosity $\lbol$ and gas radial distribution
$n(r)$, $r(\xi=10^5)$ defines the size of a Compton heating zone where
gas can achieve a steady state with $T=\tc$. Consider as a specific
example M87, a giant elliptical galaxy in the Virgo cluster. M87 hosts
a $3\times 10^9\ms$ MBH and contains hot ($T\sim 10^7$~K) gas characterized by
$n\sim$~0.1, 0.05 and 0.02 ~cm$^{-3}$ at $r=1$, 4 and 10~kpc,
respectively \citep{matetal02}. For these parameters, the outer
boundary of the Compton heating zone would be at $r(10^5)\approx
1$~kpc if the M87 nucleus switched on at its Eddington
luminosity. This size is somewhat smaller than $\rc=2.2$~kpc given by
equation (\ref{rc}) for M87.

The $\tstat(\xi)$ curve is quite different for $\xi<10^5$ in the case of the
type 2 spectrum, for almost complete lack of ionizing UV and soft
X-ray photons. In this case, there is a narrow transition region at
$\xi\sim 10^5$ dividing a high-temperature and low-temperature
regions characterized by $\tstat=\tc\sim 10^7$ and $\sim 10^4$~K,
respectively. Moreover, the solution in this transition zone
is known to be unstable \citep[e.g.][]{bufmcc74,kalmcc82}. It is 
interesting to note that in the low-temperature region, where the
gas is practically neutral, Compton heating due to scattering on
bound electrons contributes 8\% to the total heating rate dominated by
photoionization of metallic atoms by hard X-rays. This,
however, has negligible effect (less than 1\%) on the resulting
equilibrium temperature in the low-temperature region. The effect is
even smaller for the average and type 1 quasar spectra, in which case
the photoionization heating associated with the radiation above
13.6~eV is several orders of magnitude stronger than Compton heating
on bound electrons in the low-ionization limit ($\xi\lesssim 10^{-4}$).

The results presented in Fig.~\ref{xi_temp}a thus imply that
radiative heating can be quite different for $\xi\lesssim 10^5$, i.e.
outside the Compton heating zone, if plasma is exposed to the
radiation reprocessed by the obscuring torus instead of being
irradiated directly by the quasar nucleus. The direct consequence of
this is the possibility of significantly anisotropic radiative feedback on the
quasar environement in the scenario with partial obscuration of the
nucleus: strong photoionization heating (of an initially cold
material) can take place within the cones of direct nuclear emission
but not in obscured directions.

We, however, consider the above situation unlikely, because it
resulted from the complete absence of UV and soft X-ray radiation in
our adopted type 2 spectrum. In reality the heating anisotropy will be
significantly reduced if a small fraction of the ionizing radiation
outgoing from the nucleus within the open cones is scattered into
obscured lines of sight. This is demonstrated by computations we
performed for input spectra given by $\langle F_E\rangle ({\rm
type 2})+0.01\langle F_E\rangle ({\rm type 1})$ and $\langle
F_E\rangle ({\rm type 2})+0.05\langle F_E\rangle ({\rm type 1})$. This
corresponds to the scattering gas having a Thomson optical depth of
$\taut\sim 0.04$ and $\sim 0.2$, respectively, assuming that 
this gas is ionized (so that electron scattering of UV radiation is
possible) and fills the central funnel of the torus and extends some
distance above its top, covering at least $f=1/4$ of the sky as
seen from the nucleus. As can be seen from Fig.~\ref{xi_temp}a, the
addition of such a small amount of scattered ionizing radiation
changes the situation dramatically, significantly diminishing the
departure of the resulting $\tstat(\xi)$ curve from those
corresponding to the globally averaged and type 1 spectra. In such a
case, irradiation by the quasar will probably have rather similar
effects on gas located along unobscured and obscured lines of sight
not only inside the zone governed by Compton heating and cooling but
also outside it. 

Different kinds of observation do reveal significant quantities of
warm ($T\sim 10^{4.5}$--$10^6$~K) ionized gas located at distances from
possibly a fraction of a pc up to hundreds of pc from the nuclei of
Seyfert galaxies \citep[e.g.][]{krokri01}. In Seyfert 2 galaxies such
as NGC~1068, this plasma reveals itself via polarized scattered
emission in the optical \citep[e.g.][]{milgoo90} 
and via scattered continuum, recombination emission and resonance
scattering  in the X-ray band \citep[e.g.][]{turetal97,ogletal03}. The
same photoionized gas produces absorption features in the UV and soft
X-ray spectra of Seyfert 1 galaxies and some quasars
\citep[e.g.][]{geoetal98,matetal94}. Futher, $\sim 10$\% of
optically selected quasars show broad absorption lines 
(BALs), arising in material apparently flowing outward from the
nucleus with velocities $\sim 10^4$~km~s$^{-1}$, and it appears that
most or possibly all quasars contain such BAL outflows 
\citep[e.g.][]{greetal01}. Finally, powerful radio galaxies
at high redshifts show strong UV line and continuum
emission extended along the axis of the radio source, which is
interpreted as scattered radiation from an obscured quasar nucleus
\citep[e.g.][]{traetal98}. Taken together, all 
these diverse observations suggest that a warm or hot gas characterized by
the product $\taut f\lesssim$~a few per cent -- of the same order as
the scattering fractions assumed in our simulations above -- may be
ubiquitously present in AGN. We also note that electron scattering
optical depths $\taut\sim$ several $10^{-2}$ are predicted for the $\sim
100$~pc cooling flows feeding quasars in elliptical galaxies
\citep[e.g.][]{cioost01}.

We finally note that the $\tstat(\xi)$ curves shown in
Fig.~\ref{xi_temp} descend below $10^4$~K when $\xi$ becomes less than
$\sim 1$. This implies that the gas density must exceed
$10^7$~cm$^{-3}(10~{\rm pc}/r)^2$ at distance $r$ from a $10^8\ms$
black hole radiating at Eddington limit in order to provide conditions for the
existence of a dusty molecular torus such as postulated in the AGN
unification model. Such dense material cannot fill more than a small
fraction of the obscuration region, otherwise its
large Thomson optical depth $\taut>2\times 10^2(10~{\rm pc}/r)$ will
prevent even hard X-rays from escaping through the torus and make the
source Compton thick. We note that XSTAR becomes unsuitable at
$T\lesssim 3000$~K.

\section{Conclusions}
\label{concl}

The main results obtained in this work are as follows.

We combined information on the cumulative AGN light in the IR and
X-ray bands, the estimate of the local mass density of MBHs and
composite optical to soft X-ray quasar spectra to construct in a
robust way the angular-integrated radiation spectrum of 
the average quasar in the universe (\S\ref{spectrum}). This spectrum
characterizes the global energy release via accretion onto MBHs, and
is the result of implicit summing over unobscured and obscured sources.

We calculated (\S\ref{ctemp}) the Compton heating and cooling rates
for gas exposed to radiation with the adopted average quasar spectral
distribution. The Compton heating results from downscattering in
energy of the hard X-rays, while the cooling is due to inverse Compton
scattering of primarily the IR photons. The Compton temperature,
representing the equilibrium between Compton heating and cooling, is
well constrained near $2\times 10^7$~K, with an estimated uncertainty
in this value of $\sim 50$\%. 

We presented simple arguments (\S\ref{approach}) and
supported them with accurate calculations (\S\ref{ctemp}) that
circumnuclear obscuration cannot significantly affect the Compton
heating and cooling rates. As a result, the Compton temperatures
characteristic of obscured and unobscured directions (or sources) are
within a factor of 2 from the average $\tc$ value quoted above.

The almost invariant shape of the (unabsorbed) hard X-ray spectrum
combined with the recently published data of observations of tens and hundreds
of optically bright quasars that cover the blue bump segment of
the spectrum up to its connection with the X-ray component suggest that
the rms source-to-source scatter in the ratio of the amplitudes of the
hard X-ray and blue bumps is fairly small, $\sim 2$, at least for type
1 quasars. Of the same order is the possible trend going from
relatively low-luminosity ($L_X\sim 10^{44}$~erg~s$^{-1}$) to
high-luminosity ($L_X\sim 10^{46}$~erg~s$^{-1}$) quasars
(\S\ref{med}). We thus believe that the average spectrum and Compton
temperature found in this work should also represent well the spectral
output of individual quasars, at least of their majority. We note that
the IR emission from quasars results from reprocessing of the primary
UV emission, so that both components always have comparable powers,
and therefore the variability of the shape of the IR spectral
component from source to source has practically no effect on this conclusion.

Blazar-type beamed emission is energetically unimportant globally,
contributing on average less than 20\% to the net Compton heating
rate (\S\ref{blazar}). Nevertheless, current observations do not rule out the
possibility that this additional radiation component can play an
important role in gas heating for particular objects. This question
needs further study.  

We showed (\S\ref{rmax}) that during its lifetime a MBH of mass $\mbh$
can heat to the Compton temperature low-density gas within a radius of
$\sim 0.5~{\rm kpc}(\mbh/10^8\ms)^{1/2}$, which would normally
constitute a significant fraction of the effective radius of the
spheroid in which the MBH resides and a negligible fraction of the core
radius of a cluster of galaxies.

We performed computations (\S\ref{ion}) of the radiative heating of
partially photoionized plasma of cosmic chemical composition exposed
to the radiation from the average quasar, taking into account
photoionization heating and plasma cooling through line and continuum 
emission in addition to Compton heating and cooling. Although the
derived $\tstat(\xi)$ curves are quite different for the adopted
obscured and unobscured spectra, we demonstrated that scattering of a
relatively small fraction ($\sim$ a few \%) of the primary nuclear
radiation by dense gas in the vicinity of the active nucleus will wipe
away most of this difference. Observations indicate that such 
scattering does occur in most AGN. Therefore, circumnuclear
obscuration probably can be ignored to first order in considering 
the radiation feedback of MBHs on their environment.

The general results of this work provide us with the basis for follow-up
work aimed at accurately calculating the feedback effect of MBHs on
their gaseous environment.

\appendix
\section{}

In the limit $kT\ll \me c^2=511$~keV, the characteristic rates per
electron of plasma heating and cooling due to spontaneous Compton
scattering are given by \citep{sheetal88,sazsun01}
\beqa
\left(\frac{d W}{dt}\right)_{+} &=& \sigmat \me c^3
\nonumber\\
&&\int_0^\infty\epsilon_E(x)\left[\frac{3}{8x^3}(x-3)(x+1)\ln (2x+1)
\right.
\nonumber\\
&&
\left.
+\frac{-10x^4+51x^3+93x^2+51x+9}{4x^2(2x+1)^3}\right]dx
\nonumber\\
&=&
\sigmat\me c^3
\nonumber\\
&&
\int_0^\infty x\epsilon_E(x)\left[1-\frac{21x}{5}+O(x^2)\right]dx
\label{cheat}
\eeqa
and
\beqa
\left(\frac{d W}{dt}\right)_{-}
 &=& -\sigmat ckT
\nonumber\\
&&\int_0^\infty\epsilon_E(x)\left[\frac{3(3x^2-4x-13)}{16x^3}\ln(2x+1)
\right.
\nonumber\\
&&
\left.
+\frac{-216x^6+476x^5+2066x^4+2429x^3}{8x^2(2x+1)^5}
\right.
\nonumber\\
&&
\left.
+\frac{1353x^2+363x+39}{8x^2(2x+1)^5}\right]dx
\nonumber\\
&=&
-4\sigmat ckT
\nonumber\\
&&
\int_0^\infty\epsilon_E(x)\left[1-\frac{47x}{8}+O(x^2)\right]dx,
\label{ccool}
\eeqa
where $x=E/\me c^2$ and $\epsilon_E(x)$ is the radiation spectral
energy density. It follows that $a(E)=1-21E/5\me c^2+...$ and
$b(E)=1-47E/8\me c^2+...$ in equations (\ref{tc1}) and (\ref{tc2}). 


\end{document}